\documentclass[preprintnumbers, twocolumn, a4paper,prb]{revtex4}
\usepackage{graphicx}
\usepackage{graphics}
\usepackage{amsmath, amsfonts, amssymb, bm}

\usepackage{amsmath}
\usepackage{verbatim}
\usepackage{bm}
\usepackage{cases}
\usepackage{color} %for colored fonts
\usepackage{epsfig}

\begin{document}

\title{Experimental realisation of tunable ferroelectric/superconductor $(\text{B}\text{T}\text{O}/\text{Y}\text{B}\text{C}\text{O})_{\text{N}}/\text{S}\text{T}\text{O}$ 1D photonic crystals in the whole visible spectrum} 

\author{Luz~E. Gonz\'alez$^{1,2,3}$}
\email{luz.gonzalez@correounivalle.edu.co}
\author{John~E. Ordo\~nez$^{4}$}
\email{john.ordonez@correounivalle.edu.co}
\author{Carlos~A.~Melo-Luna$^{2,5}$}
\email{carlos.melo@correounivalle.edu.co}
\author{Evelyn~Mendoza$^{4}$}
\author{David~Reyes$^{6}$}
%\author[5]{J.~Douin}
\author{Gustavo~Zambrano$^{4}$}
\author{Nelson~Porras-Montenegro$^{1,2}$}
\author{Juan~C.~Granada$^{1,2}$}
\author{Maria~E.~G\'omez$^{4}$}
\author{John~H.~Reina$^{2,5}$}
\email{john.reina@correounivalle.edu.co}
\affiliation{$^{1}$Solid State Theoretical Physics Group, 
Departamento de F\'isica, Universidad del Valle, 760032 Cali, Colombia}
\affiliation{$^{2}$Centre for Bioinformatics and Photonics (CIBioFi) and Departamento de F\'i­sica, Universidad del Valle, Edificio E20~No.~1069, 760032 Cali, Colombia}
\affiliation{$^{3}$Facultad de Ciencias Naturales y Matem\'aticas, Universidad de Ibagu\'e, 730001 Ibagu\'e, Colombia}
\affiliation{$^{4}$Thin Films Group, 
Departamento de F\'i­sica, Universidad del Valle, 760032 Cali, Colombia}
\affiliation{$^{5}$Quantum Technologies, Information and Complexity Group, Departamento de F\'i­sica, Universidad del Valle, 760032 Cali, Colombia}
\affiliation{$^{6}$Centre d'\'E‰laboration de Mat\'eriaux et d'Etudes Structurales (CEMES) CNRS-UPR 8011, 29 rue Jeanne Marvig, 31055 Toulouse, France}
%%%%%%%

\begin{abstract}
{Emergent technologies that make use of novel materials and quantum properties of light states are at the forefront in the race for the physical implementation,  encoding and transmission of information. Photonic crystals (PCs) enter this paradigm with optical materials that allow the control of light propagation and can be used for optical communication~\cite{Shalaev:2019}, and photonics and electronics integration~\cite{Marpaung:2019, Gurioli:2019} making use of materials ranging from semiconductors~\cite{Priolo:2014}, to metals~\cite{Pater:2018}, metamaterials~\cite{Segal:2015}, and topological insulators~\cite{Yang:2019}, to mention but a few. In particular, here we show how designer superconductor materials integrated into PCs fabrication allow for an extraordinary reduction of electromagnetic waves damping and possibilitate their optimal propagation and tuning through the structure, below critical superconductor temperature. We experimentally demonstrate, for the first time, a successful integration of ferroelectric and superconductor materials into a one-dimensional (1D) PC composed of $(\text{B}\text{T}\text{O}/\text{Y}\text{B}\text{C}\text{O})_{\text{N}}/\text{S}\text{T}\text{O}$ bilayers that work in the whole visible spectrum, and below (and above) critical superconductor temperature (measured in the 10-300 K range). Theoretical calculations support, for different number of bilayers $N$, the effectiveness of the produced 1D PCs and pave the way for novel optoelectronics integration and information processing in the visible spectrum at low temperature, while preserving their electric and optical properties.}
\end{abstract}

\maketitle
%\keywords{photonic crystal,visible light radiation,superconductor,ferroelectric,heterostructure,laser excitation}
\section{introduction}
The use of electromagnetic (EM) waves as information carriers for communication systems has been in place for many years~\cite{Bell:1880,Andrews:2001}; as such, EM wavelengths make possible the transmission over large distances but, at the same time, they limit the amount of information they can convey by their frequency: the larger the carrier frequency, the larger the available transmission bandwidth and thus the information-carrying capacity of the communication system~\cite{Senior:2009}. For this reason, quantum artificial nanostructured materials such as photonic crystals that are able to transmit at high frequencies and that concentrate the available power within the transmitted electromagnetic wave, thus giving an improved system performance, are desired~\cite{Joannop:2009}. PCs are artificial periodic structures characterised by a periodic variation of the refractive index with a consequent periodic spatial variation of the dielectric constant, which may be tailored to control light properties~\cite{Joannop:2009}. They, therefore, allow the appearance of defined frequency ranges and address the issue of forbidden/allowed propagation of electromagnetic waves. As a consequence, the control and tunability of PCs opens a new perspective for information processing and technological applications such as chips~\cite{Wang:2004, Pan:2010}, filters~\cite{Wang:2019}, lasers~\cite{Xue:2016}, waveguides~\cite{Park:2008, Susa:2014}, integrated photonic circuits~\cite{Kuram:2014}, chemical and biological sensors~\cite{Boris:2015}, and thin film photovoltaics~\cite{Nirmal:2017, Madrid:2019}, to cite but a few. 

Even though there exist many sort of materials used for tunable PCs~\cite{Priolo:2014, Pater:2018, Segal:2015, Yang:2019}, substantial advances would be expected if superconductor  properties (YBCO) could be merged with those of ferroelectric ones (BTO), on a photonic structure. As it is well known, superconductors are materials characterised for low losses and better operating characteristics than normal metals. Nowadays, they are considered as promising quantum materials, widely used in quantum computing networks~\cite{Basov:2017,Tokura:2017,Gambetta:2017, Melo-Luna:2017}, quantum simulators~\cite{ Houck:2012,Lamata:2017},  loss-less microwave resonators~\cite{ Hua:2016},  and AC Josephson junction lasers~\cite{ Welp:2013}. Superconductors used as building blocks for PCs fabrication have mainly two advantages over traditional ones. First, the damping (losses)-issue of electromagnetic waves present in metals, can be overcome by utilisation of superconductors instead. Second, the dielectric function of the superconductor is mainly dependent on the London penetration depth, which is a function of temperature and external magnetic field~\cite{Ooi:2000, Takeda:2003, Berman:2006}. On the other hand, ferroelectric thin films technological features of great significance such as short response time~\cite{Martin:2016}, remanent polarization~\cite{Rault:2014}, faster tuning compared to ferromagnetic materials~\cite{Boyn:2017}, smaller and lighter structures~\cite{Wang:2018}, high power capacity~\cite{Cheng:2017}, and Pockels modulation~\cite{Koen:2018}, make of the BTO the subject of a number of investigations for developing high-performance electro-optical modulators~\cite{Castera:2015,Rosa:2017}, photodetectors~\cite{Nan:2017}, and novel devices such as, non-volatile optical memories~\cite{Xi:2017,Jieji:2016}. In particular, the latter could serve, in a wide range of applications, as a high-speed optical buffer memory by means of the optical fiber high broadband, with a significant reduction of power consumption for data processing. 

Prompted by the above grounds, here we present a robust integration of ferroelectric/superconductor materials in a 1D photonic heterostructure with remarkable optical properties that work at low (below $T_C$) and above $T_C=80$ K,  which can be harnessed for optical communication and optoelectronics integration. We have realised and experimentally demonstrated the optical reliability of the fabricated nanosystem via reflectance spectrometry measurements in the whole visible range, and also theoretically modelled below and above superconductor critical temperature for small and large number of heterostructure periods, bilayers $N$. Our findings pave a way for novel PC materials development that can be merged into photonic integrated circuits, optoelectronics, or in devices for the transmission of information in the visible range at low temperature environments, while preserving their electric and optical properties.

\section{Results and Discussion}
\subsection{Structural properties}
%%%%%%%%%%%%%%%%%%%%%%%%%%
\begin{figure*}[t]
%\begin{figure}[h]
\centering
\includegraphics[width=1.0\linewidth]{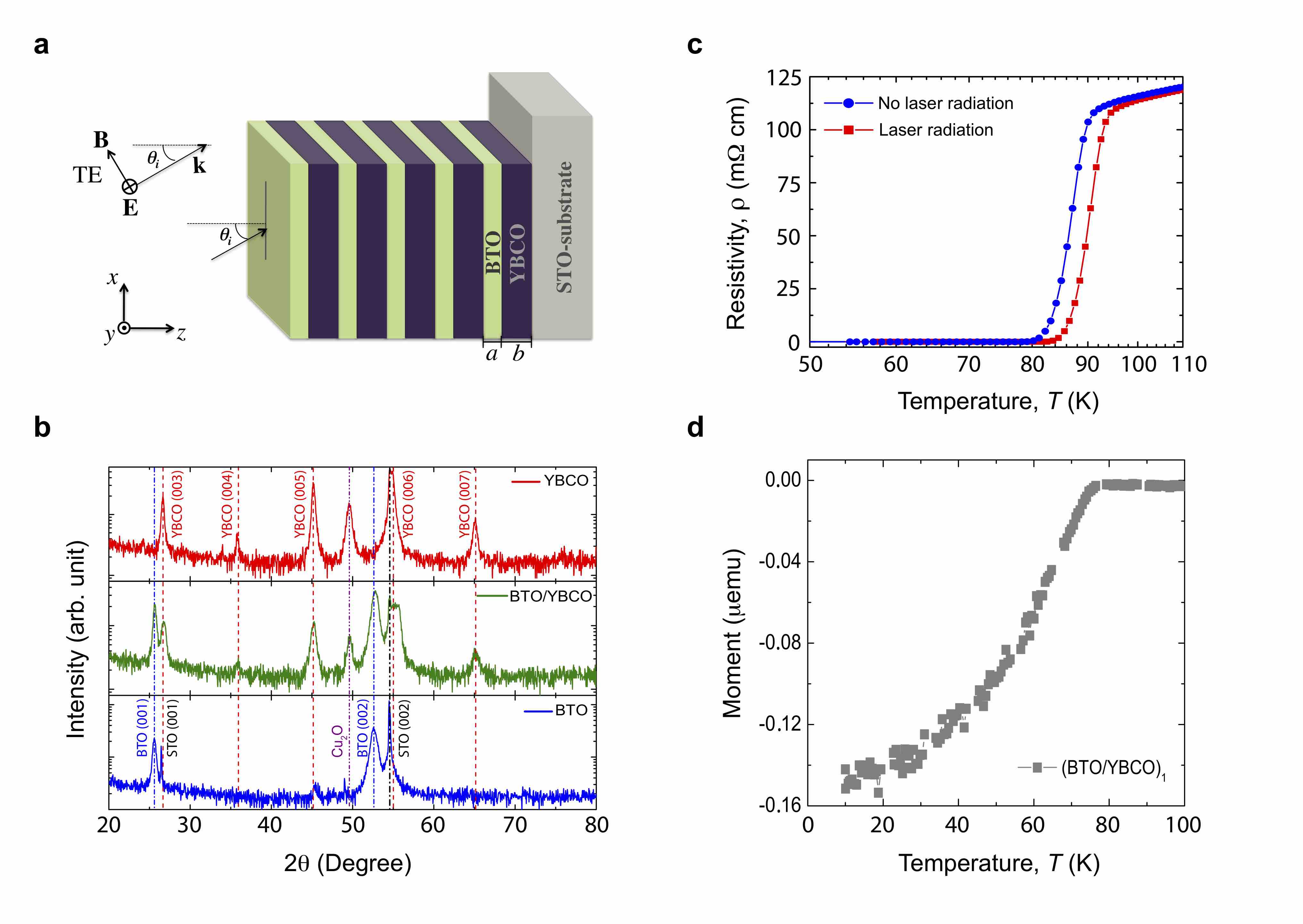}
\caption{\textbf{a}, Schematic diagram of $(\text{B}\text{T}\text{O}/\text{Y}\text{B}\text{C}\text{O})_{\text{N}}/\text{S}\text{T}\text{O}$ 1D PC for $N = 1, 3, 5$ periods. The films were fabricated by RF sputtering onto a polished $\text{Sr}\text{Ti}\text{O}_{3}$ (001) substrate. YBCO and BTO thicknesses correspond to $b=73$ nm and $a=30$ nm, respectively. \textbf{b}, X-ray $\theta-2\theta$ scans for YBCO(70 nm)/STO film, BTO(30 nm)/YBCO(73 nm)/STO bilayer, and BTO(30 nm)/STO film. The dashed vertical lines are associated to pure phases.  A small amount of Cu$_{2}$O phase is identified for the YBCO layer. \textbf{c}, Resistivity as a function of temperature for an YBCO film (blue dots) and for the same YBCO film under a coherent laser radiation (red squares). \textbf{d}, Magnetization with temperature dependence at zero field cooling (ZFC) with an applied field  $H$ = 1 kOe for BTO(30nm)/YBCO(73 nm)/STO for $N=1$ (green squares).}
\label{fig1}
\end{figure*}
%%%%%%%%%%%%%%%%%%%%%%%%%%
%%%%%%%%%%%%%%%%%%%%%%%%%%%%%%%%%%%%%%
\begin{figure*}[t]
\centering
\includegraphics[width=1.0\linewidth]{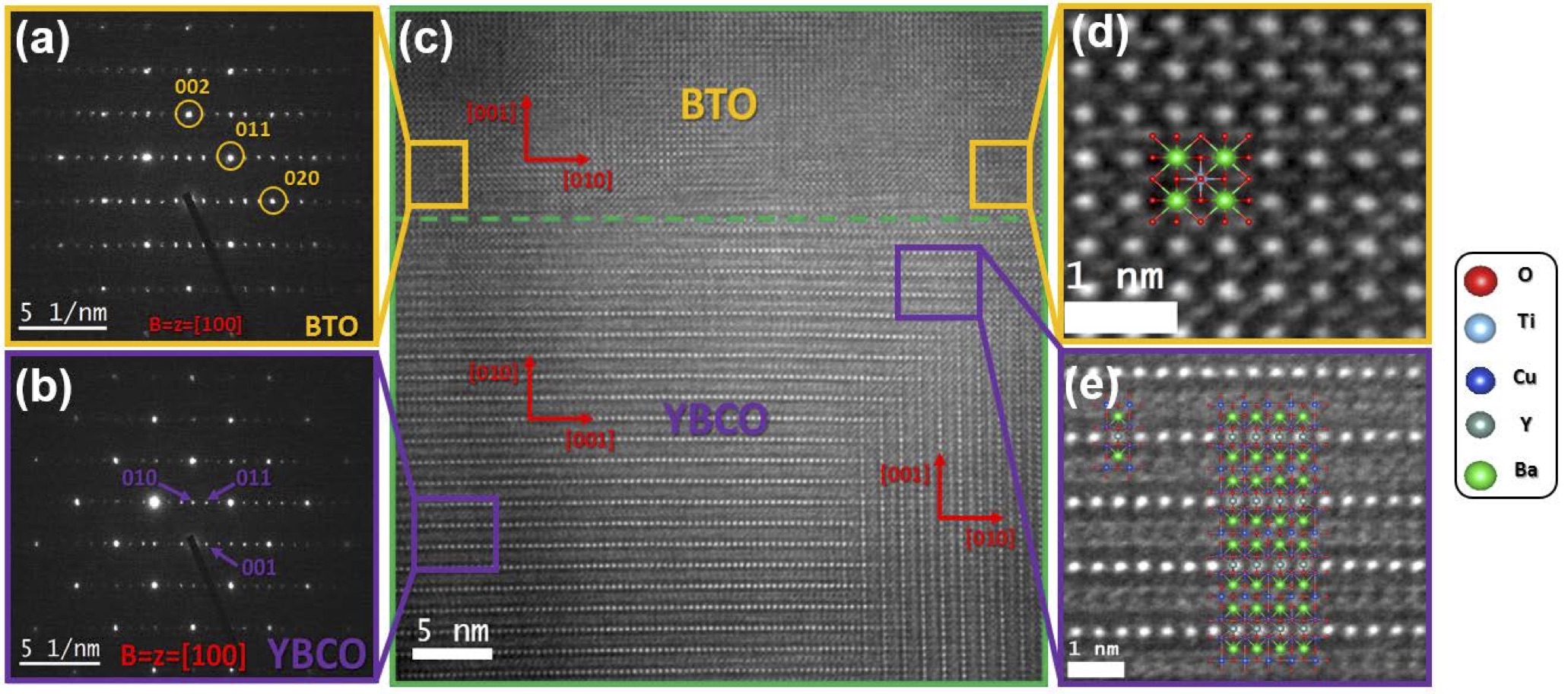}
\caption{Electron diffraction pattern to \textbf{a} YBCO and \textbf{b} BTO layers with different crystal orientations are displayed. \textbf{c}, TEM image for (BTO/YBCO)$_{N=1}$ with associated [001] and [010] crystal directions. Zoom of HRTEM images for the YBCO \textbf{d} and BTO \textbf{e}, respectively. A sketch of the crystal structure displays the atomic positions.}
\label{fig2}
\end{figure*}
%%%%%%%%%%%%%%%%%%%%%%%%%%%%%%%%%%%%%%

Figure~\ref{fig1}a shows a schematic diagram of three 1D photonic crystals of one-, three-, and five-pair of BTO/YBCO bilayers, fabricated by DC and RF sputtering onto polished $\text{Sr}\text{Ti}\text{O}_{3}$ (001) substrates. Here, $a$ and $b$ correspond to the thicknesses of $\text{Ba}\text{Ti}\text{O}_{3}$ and $\text{Y}\text{Ba}_{2}\text{Cu}_{3}\text{O}_{7}$, $\theta_{i}$ denotes the angle with the {\it z}-axis defined in the range of $0^{\circ} \sim \pm 90^{\circ}$, {\it xz} is the plane of incidence, and the direction of $\textbf{E} \times \textbf{B}$ is given by the incident wave vector \textbf{k}, where \textbf{E} and \textbf{B} represent the electric and magnetic fields, respectively. Figure~\ref{fig1}b displays the out-of-plane XRD $\theta$-$2\theta$ scans for the YBCO(70 nm)/STO film, the BTO(30 nm)/YBCO(73 nm)/STO bilayer, and the BTO(30 nm)/STO film. YBCO and BTO peaks associated to the (001) direction were identified for reflections from 20$^{\circ}$ to 80$^{\circ}$. The peaks observed in the individual layers are indexed in the bilayer, and indicate a textured growth of both samples. A minor $\text{Cu}_{2}\text{O}$ phase was identified in our samples; this impurity is typical for YBCO ~\cite{Porat:1994,Chen:2016}. We obtained the following lattice parameters: $\textit{a}_{BTO}=4.035$ $\textup{\AA}$, $\textit{a}_{YBCO}= 3.878$ $\textup{\AA}$ in films, while $\textit{a}_{BTO}=4.040$ $\textup{\AA}$ and $\textit{a}_{YBCO}= 3.867$ $\textup{\AA}$ for the BTO/YBCO bilayer. For a bilayer, not considerable displacement of peaks was observed compared to YBCO and BTO films (dashed vertical lines) grown under identical parameters. Hence, there is no effect due to the BTO layer on the position of the Bragg reflection peaks for the YBCO layer.

In our case, the YBCO superconductor state corroboration is necessary to confirm the SC/dielectric PC formation. Thus, in fig.~\ref{fig1}c, a traditional resistance with temperature dependence is observed (blue dots), where a superconducting temperature T$_{SC}$ $\sim$ 85 K to YBCO/BTO film was identified. However, during (BTO/YBCO)$_{N}$-PC structure measurements, the YBCO layer was exposed to laser radiation at different wavelengths and the possibility of a breakdown of the Cooper pairs could occur during the radiation-matter interaction (e.g., in YBCO intergranular films the radiation damage requires  $\sim$ 1 eV/atom)~\cite{Chrisey:1989,Wei:1991,Quere:1988}. In our case, for the YBCO/STO film, a resistance measurement was performed while the sample was exposed to a coherent laser radiation source in the 500 nm - 800 nm range (red squares), as shown in fig.~\ref{fig1}c. For YBCO, the superconducting gap equals 30 meV, and the laser radiation energy which would break down the Cooper pairs is far away from this, in the range between 2.5 eV (800 nm) and 3.0 eV (400 nm)~\cite{Poole:2007}. Furthermore, the sample is irradiated with a power below 10 mW, hence the sample remains in a superconducting state, even though the energy of the laser radiation is a hundred times greater than the superconducting gap. However, only a slight displacement associated to thermal effect was found when the laser is on and T$_{SC}$ ($\sim$ 87 K), i.e., in our case, the radiation-matter interaction does not significantly influence the YBCO superconductivity and allows for a PC in the superconductor state. For $(\text{B}\text{T}\text{O}/\text{Y}\text{B}\text{C}\text{O})_{N}$ multilayers, the superconductivity state is indirectly measured from YBCO diamagnetism~\cite{Springer:2016}. In fig.~\ref{fig1}d, the temperature-dependent magnetisation for the $(\text{B}\text{T}\text{O}/\text{Y}\text{B}\text{C}\text{O})_{1}$ multilayer is depicted, and we reach a superconducting transition temperature decrease to T$_{SC}$  $\sim$ 70 K. This decrease can be associated to different mechanisms such as tensile/compressive strain in BTO/YBCO interfaces, interface atom migration during the film growth~\cite{Alberca:2015}, oxygen losses~\cite{Manthiram:1987,Bruynseraede:1989}, dislocations~\cite{Pan:2003}, Cooper pairs breaks~\cite{Zhang:2016,Gedik:2004,Yang:2015}, among others~\cite{Chisholm:1991}, but we stress that the superconducting state in the PC is preserved.

We analysed the BTO/YBCO interface by means of transmission electron microscopy (TEM). Initially, an electron diffraction pattern for the YBCO and BTO layers are presented in fig.~\ref{fig2}a and b (the electron diffraction patterns were rotated 90$^{\circ}$ in order to match with the high resolution images orientations). The lamella was slightly tilted in order to achieve the closest zone axis for the STO crystal. In Fig.~\ref{fig2}c, a high resolution image shows the YBCO/BTO bilayer with a flat interface and the layers orientation. As can be seen, the YBCO layer mainly grows in the [010] direction, but also a few grains growth along the [001] orientation were found. Additionally, the BTO growth along the [001] orientation is in agreement with the STO substrate growth direction. Finally, in fig.~\ref{fig2}d and e, a high-resolution image for YBCO and BTO, respectively, are displayed, where the atomic columns are visible for some of the atoms. The arrangement of the different atoms is represented by the unit cell for BTO (Fig.~\ref{fig2}d) and YBCO (Fig.~\ref{fig2}e) layers. In the case of YBCO, a supercell composed by 16 unit cells shows in a better way the match between the crystal structure and the high resolution image.

\subsection{Optical response: Theory and Experiment}

%%%%%%%%%%%%%%%%%%%%%%%%%%%%%%%%%%%%%%
\begin{figure*}[t]
\centering
\includegraphics[width=1.0\linewidth]{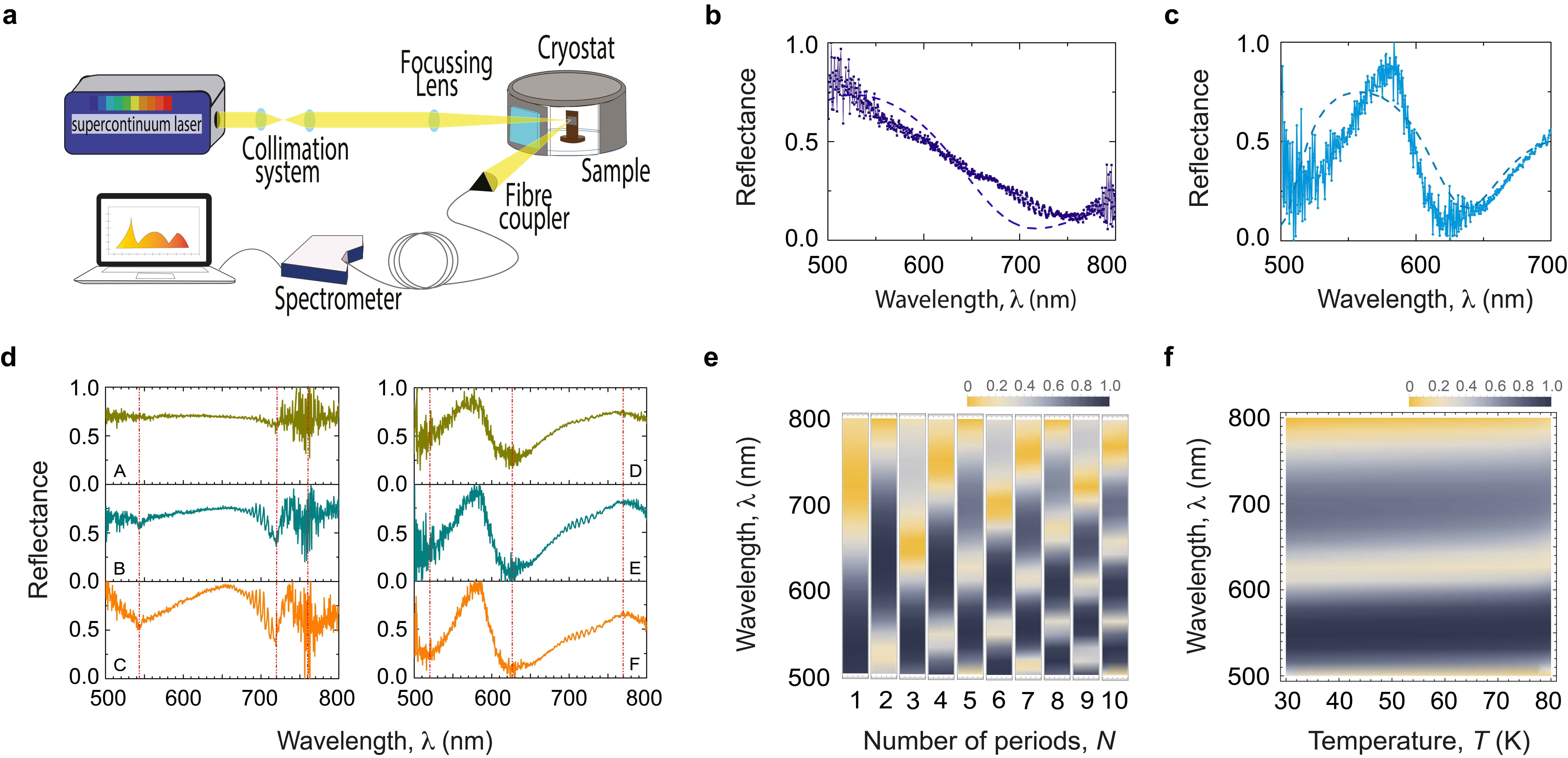}
\caption{\textbf{a}, Optical setup. A supercontinuum laser source in the wavelength range between 400-800 nm, collimated with a Galilean telescope lenses setup, is sent to the multilayer $(\text{B}\text{T}\text{O}/\text{Y}\text{B}\text{C}\text{O})_{\text{N}}$ sample inside a cryostat, at a controllable temperature, toward mirrors and a focusing lens. The reflection from the sample is collected by a fiber coupler into a multimode fiber towards the spectrometer and the response is analysed in a workstation. \textbf{b} and \textbf{c}, Reflectance response of $(\text{B}\text{T}\text{O}/\text{Y}\text{B}\text{C}\text{O})_{\text{N}}/\text{S}\text{T}\text{O}$ 1D PC for $N=1$, and 5, respectively. Continuous and dashed curves correspond to experimental and theoretical results at $T=50$ K, respectively. \textbf{d}, Reflectance response of $(\text{B}\text{T}\text{O}/\text{Y}\text{B}\text{C}\text{O})_{\text{N}}/\text{S}\text{T}\text{O}$ 1D PC for (A) $T=80$ K, (B) $T=50$ K, and (C) $T=30$ K, for $N = 3$, and panels (D) $T=80$ K, (E) $T=50$ K, and (F) $T=30$ K, for $N = 5$. \textbf{e}, Simulated gap map of the $(\text{B}\text{T}\text{O}/\text{Y}\text{B}\text{C}\text{O})_{N}/\text{S}\text{T}\text{O}$ 1D PC as a function of wavelength and number of periods $N$. The width of each bar corresponds to a temperature range between $50$ and 60 K. \textbf{f}, Simulated gap map of the $(\text{B}\text{T}\text{O}/\text{Y}\text{B}\text{C}\text{O})_{5}/\text{S}\text{T}\text{O}$ 1D PC as a function of wavelength and temperature $T$, from $T=30$ K to 80 K. The dark areas correspond to the photonic band gaps or high-reflectance ranges and yellow regions indicate high transmission bands where radiation passes through the structure.}
\label{fig3}
\end{figure*}
%%%%%%%%%%%%%%%%%%%%%%%%%%%%%%%%%%%%%%

Figure~\ref{fig3}a shows the schematics of the experimental setup used to measure the optical response of the  $(\text{B}\text{T}\text{O}/\text{Y}\text{B}\text{C}\text{O})_{N}/\text{S}\text{T}\text{O}$ 1D PC (described in Methods section). In Fig.~\ref{fig3}b and c, we display  the effect of the number of periods on the reflectance spectra. Continuous and dashed curves correspond to the experimental and theoretical results obtained when $N$ is equal to 1 and 5, and the temperature is kept constant at $T=50$ K, respectively.  Even though  measured spectra are in agreement with the theoretical predictions, the overall behaviour presents a considerable number of resonant peaks and a small variation in reflectance intensity. Based on the TEM analysis of Fig.~\ref{fig2}, we associate these peaks to the possible nonuniformity of the thickness and the additional presence of few YBCO grains grown along the [001] direction. Additionally, a correspondence between the number of interference fringes (or bands) and the number of periods of bilayers in the structure is noticeable. We argue that these fringes are the result of the interference of incident light beams partially reflected and transmitted at the interfaces between layers.
 
To gain further insight into the optical response of our structure we used the transfer matrix method (see the Methods section) to calculate the reflectance spectra ~\cite{Gonzalez:2017, Markos:2008}, and the two fluid model~\cite{Orlando:1991}, to consider the contribution of the superconductor (YBCO) to the dielectric response of the PC. As we describe below, this model allowed us to theoretically explain the reflectance experimentally found. In order to examine in detail the bands dependence on the number of periods, in Fig.~\ref{fig3}e we plot the simulated gap map in the whole range from $N = 1$ to 10. The dark areas correspond to the photonic band gaps (PBGs) or high-reflectance ranges, while yellow areas indicate transmission ranges where radiation passes trough the structure. It is noticeable for $N = 1$ that the reflectance decreases from 500 nm to approximately 610 nm (see Fig.~\ref{fig3}b), which is observed in the gap map of Fig.~\ref{fig3}e as the change from dark to yellow region. In the case $N = 5$, two ranges of low reflectance are seen around 510 nm and 620 nm (Fig.~\ref{fig3}c), which is completely in agreement with the theory. We have extended our results up to $N = 10$, to show the sensitivity of the PBGs with $N$: the larger $N$,  the lower the wavelengths of the photonic bandgaps, and  their gap structure experiences a switch for the EM waves propagation from forbidden to allowed frequency ranges at a given wavelength. Interestingly, such structured gaps become narrower as $N$ changes from an even to an odd number, which could be exploited as an optical switch.

The two-fluid model is often used to describe the behaviour of a superconductor at nonzero temperature. It consists of two distinct noninteracting fluids of electrons that carry current, where each fluid follows two parallel channels, one superconductor and one normal. Accordingly, when a material is superconducting, some of the electrons will be superconductors and some will still be normal electrons. Thus, there will be a mixture of superelectrons and normal electrons. For these reason, we can model the total conductivity as follows: for $T \leq T_{c}$, as the sum of the normal conductivity maintained by unpaired electrons, and the superconducting conductivity maintained by superelectrons; and for $T > T_{c}$, by the Drude conductivity for a normal metal. More explicitly, it can be written as~\cite{Orlando:1991}: 

 %%%%%%%%%%%
 \begin{subnumcases}{\sigma_{s}=} 
              \frac{i}{\omega\mu_{0} {\lambda_{L}}^{2}(T)}+ \frac{nq^{2}\tau}{m}\frac{1}{1-i\omega\tau}f_{n}, &      $T \leq T_{c}$ \label{eq1a}
              %\\ \nonumber
             \\ \frac{nq^{2}\tau}{m}\frac{1}{1-i\omega\tau}, &     $T > T_{c}$  \label{eq1b}
 \end{subnumcases}
 %%%%%%%%%%%%
 where $\omega$ is the EM wave frequency, $\mu_{0}$ is the permeability of free space; $n$, $q$, and $m$ are respectively the density,  the charge,  and the mass of the carrriers, $\tau$ denotes the scattering time of electrons. The damping frequency $\gamma=\frac{1}{\tau}$, $f_{n}={ \left( \frac { T }{ { T }_{ c } }  \right)  }^{ p }$ gives the density of normal state electrons over the total number of electrons, and $\lambda_{L}$ is the temperature-dependent penetration depth~\cite{Orlando:1991}:
%%%%%%%%%%%%%%
\begin{equation}
\label{eq2}
{ \lambda  }_{ L }=
\frac { { \lambda  }_{ 0 } }{ \sqrt { 1-{ \left( \frac { T }{ { T }_{ c } }  \right)  }^{ p } } },
\end{equation}
where $\lambda_{ 0 }$ is the value of the penetration depth at zero temperature, and the exponent $p$ corresponds to 2 and 4 for high and low temperature superconductors, respectively. 

The dielectric response of a given material is introduced by means of the electric permittivity. This parameter is in general a complex number, $\epsilon=\epsilon_{r}+i\epsilon_{i}$, where the imaginary part $\epsilon_{i}$ accounts for electromagnetic losses in the material and is closely related to the current inside the material through the complex conductivity $\sigma(\omega)$, such that
$\epsilon = \epsilon(\omega) =  \epsilon_{\infty}+i \frac{\sigma(\omega)}{\epsilon_{0}\omega}$, 
where $\epsilon_{\infty}$ is the dielectric function at high frequencies and $\epsilon_{0}$ the permittivity of free space~\cite{Bennemann:2008}. By replacing equation \eqref{eq1a} into $\epsilon(\omega)$, we arrive at the dielectric function of the superconductor
\begin{equation}
\label{eq4}
\epsilon(\omega, T) =  \epsilon_{\infty}-\frac{c^{2}}{\omega^{2}{\lambda_{L}}^{2}(T)}- \frac{{{\omega_{p}}^{2}}{\tau^{2}}}{1+{\omega^{2}}{\tau^{2}}}f_{n}+i \frac{{\omega_{p}}^{2}\tau}{\omega (1+{\omega^{2}}{\tau^{2}})}f_{n}.
\end{equation} 
Finally, substitution of equation \eqref{eq1b} leads to the dielectric function of a metal
\begin{equation}
\label{eq5}
\epsilon(\omega) =  \epsilon_{\infty}-\frac{{\omega_{p}}^{2}}{\gamma^{2}+\omega^{2}}+i \frac{{\omega_{p}}^{2}\gamma}{\omega(\gamma^{2}+\omega^{2})}.
\end{equation} 

To calculate the reflectance spectra, in our numerical simulations layer 1 corresponds to $\text{Ba}\text{Ti}\text{O}_{3}$ with $a=30$ nm; $\epsilon=5.8$~\cite{Wemple:1968}. As the operation temperature in the present experiment varies from 20 K to 150 K, the $\text{Ba}\text{Ti}\text{O}_{3}$-dielectric constant can be considered constant in this range~\cite{Kay:1949,Merz:1949}. Layer 2 corresponds to $\text{Y}\text{Ba}_{2}\text{Cu}_{3}\text{O}_{7}$, with $p=2$, $T_{c}=80$ K (the critical temperature experimentally obtained in this work), $b=73$ nm, the plasma frequency $\omega_{p}=1.7 \times 10^{15}$ rad/s~\cite{Lin:2010}, the damping frequency $\gamma=1.3 \times 10^{13}$ rad/s~\cite{Lin:2010}, and the dielectric constant is modelled by equation  \eqref{eq4}. According to the growth direction of the bilayers (c-axis), we consider the parallel propagation to the c-axis of $\text{Y}\text{B}\text{C}\text{O}$, i.e., the magnetic field $\vec{H}$ perpendicular to $\hat{c}$ (TE case). Therefore, the penetration depth is on the a-b plane with a $\lambda _{\perp 0}=118.6$ nm~\cite{Kamal:1994}. For temperatures close to the critical temperature $T_{c}$, over the range $0.001 < \frac{T_{c}-T}{T}< 0.1$, we adopted the magnetic penetration depth ${ \lambda  }_{ L }=\lambda _{\perp 0}[1-(T/T_{c})]^{p}$, with $p=-\frac{1}{3}$~\cite{Kamal:1994}. Simultaneously, we considered the thermal expansion effect on $\text{Ba}\text{Ti}\text{O}_{3}$ and $\text{Y}\text{Ba}_{2}\text{Cu}_{3}\text{O}_{7}$ thicknesses. In certain temperature ranges, the thermal expansion effect adopts the law $d(T)=d_{0}(1+\alpha\Delta T)$, where $\alpha$ is the thermal expansion coefficient, $\Delta{T}$ is the temperature deviation, and $d$ and $d_{0}$ are the thicknesses of each layer under actual and room temperature, respectively~\cite{Kawashima:1998}. We consider the thermal expansion coefficients to be $7.14 \times 10^{-6} /^{\circ}$C and $13.4 \times 10^{-6} /^{\circ}$C for $\text{Ba}\text{Ti}\text{O}_{3}$~\cite{Keyston:1959} and $\text{Y}\text{Ba}_{2}\text{Cu}_{3}\text{O}_{7}$~\cite{Wu:2015} , respectively.

In a previous work~\cite{Gonzalez:2017}, we have theoretically studied the transmission of the superconductor PC as a function of the wavelength for different temperatures. We found no noticeable changes in the transmission spectra with temperature, but the existence of small shifts in the PBGs were observed. With the aim to compare with experimental results, in Fig.~\ref{fig3}d, we plot in six panels, the temperature effect on the measured reflectance spectra. From panels (A), (B), and (C), where the temperature was varied from 80 K, 50K to 30 K for $N = 3$, an almost negligible shift is perceived in the wavelength ranges where the PBGs are present. Similar features were obtained in panels (D), (E) and (F), for $N = 5$ with $T=80$K, 50K and 30 K, respectively. In figure ~\ref{fig3}d we have indicated with vertical red lines specific wavelengths ($\lambda= 543$, 720, and 760 nm for $N=3$, and $\lambda= 520$, 625, and 770 nm for $N=5$) to get a better visualisation of these findings. In addition, in Fig.~\ref{fig3}f, we show the $(\text{B}\text{T}\text{O}/\text{Y}\text{B}\text{C}\text{O})_{5}/\text{S}\text{T}\text{O}$ 1D PC reflectance spectrum as a function of wavelength and temperature $T$, from $T=30$ K to 80 K, below the critical temperature of the superconductor. As in the previous case, dark areas correspond to the photonic band gaps (PBGs) or high-reflectance ranges, while yellow areas, indicate high transmission ranges where radiation passes trough the structure. We obtain a negligible shift as the temperature increases. We associate our findings with the slight decreasing suffered by the superconductor dielectric constant as the wavelength increases. If this change were appreciable, the reflectance should suffer a displacement, in agreement with the electromagnetic variational principle~\cite{Joannop:2009}.  Thus, our results demonstrate the effectiveness of the implemented $(\text{B}\text{T}\text{O}/\text{Y}\text{B}\text{C}\text{O})_{N}/\text{S}\text{T}\text{O}$ 1D PCs as an excellent candidate for the design of an optical transmitter/reflector below and above critical superconductor temperature. The fact that temperature does not significantly affect the operation frequencies of the PBGs becomes an advantage since  high reflectances can be achieved in that whole range of temperatures.

%%%%%%%%%%%%%%%%%%%%%%%%%%%%%%%%%
\begin{figure*}[t]
\centering
\includegraphics[width=1.0\linewidth]{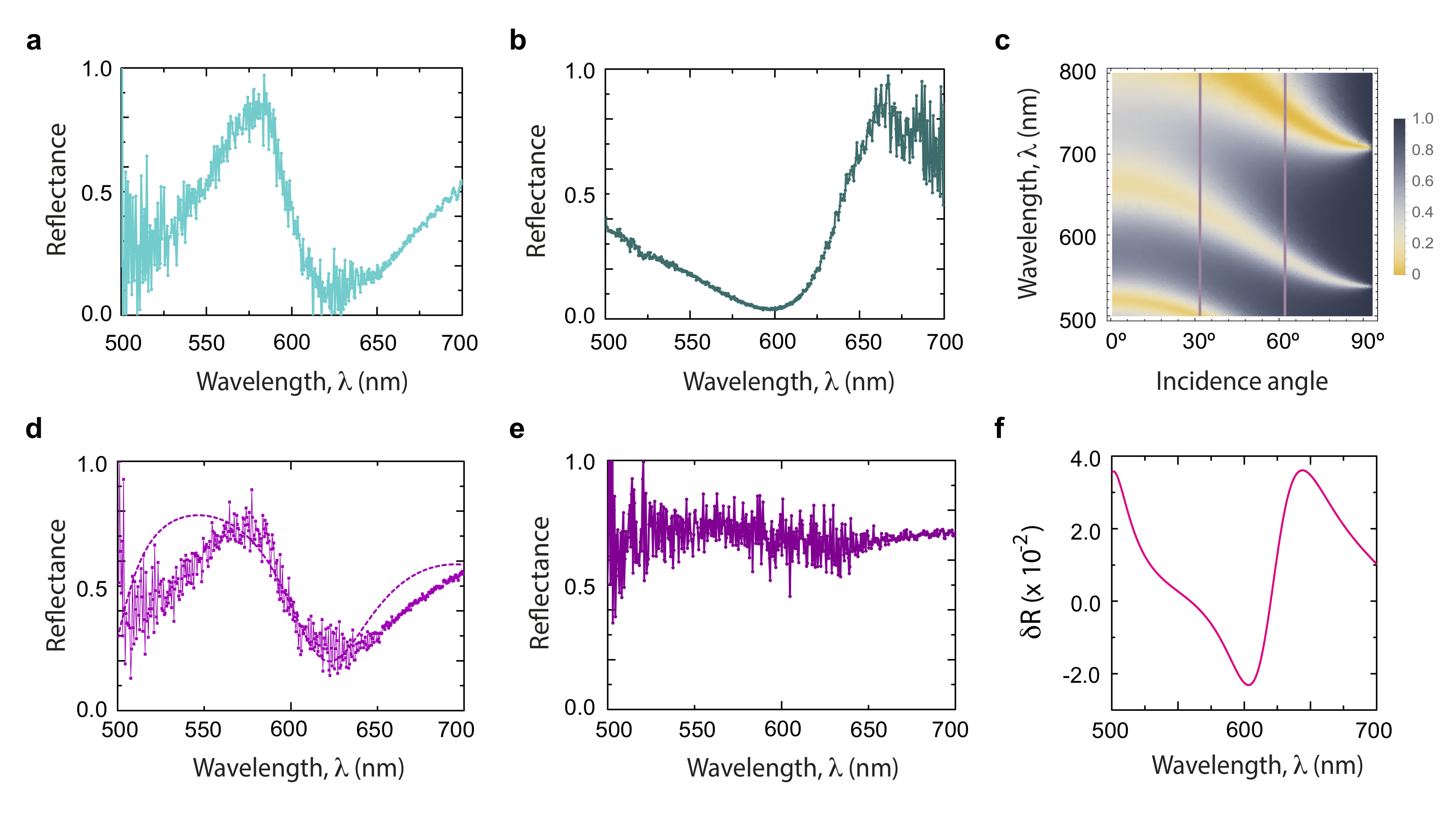}
\caption{Panels \textbf{a} and \textbf{b} display the measured reflectance of $(\text{B}\text{T}\text{O}/\text{Y}\text{B}\text{C}\text{O})_{5}/\text{S}\text{T}\text{O}$ 1D PC at $T= 50$ K and incident angles of $35^{\circ}$ and $65^{\circ}$ for TE polarization, respectively. \textbf{c}, Shows the projected gap map for TE polarization in the whole range of incident angles also at $T= 50$ K. The dark areas correspond to the photonic band gaps or high-reflectance ranges and yellow regions indicate high transmission bands where radiation passes through the structure.  Magenta vertical lines in the figure, are a guide to the eye, and correspond to the results for $35^{\circ}$ (Fig.~\ref{fig4}a) and $65^{\circ}$ (Fig.~\ref{fig4}b), respectively. \textbf{d}, Experimental (continuous curve) and theoretical (dashed curve) reflectance spectra for $(\text{B}\text{T}\text{O}/\text{Y}\text{B}\text{C}\text{O})_{5}/\text{S}\text{T}\text{O}$ 1D PC at $T=80$ K, show the agreement between experiment and the theory based on the Drude model (eq. \eqref{eq5}). \textbf{e}, $(\text{B}\text{T}\text{O}/\text{Y}\text{B}\text{C}\text{O})_{5}/\text{S}\text{T}\text{O}$ experimental spectrum at $T=100$ K, above superconductor critical temperature $T_{c}$. \textbf{f}, Contribution due to superconductor carriers on the optical response of the $(\text{B}\text{T}\text{O}/\text{Y}\text{B}\text{C}\text{O})_{5}/\text{S}\text{T}\text{O}$ 1D PC.}
\label{fig4}
\end{figure*}
%%%%%%%%%%%%%%%%%%%%%%%%%%%%%%%%%

Figure~\ref{fig4}a and b plot the measured reflectance for $(\text{B}\text{T}\text{O}/\text{Y}\text{B}\text{C}\text{O})_{5}/\text{S}\text{T}\text{O}$ 1D PC in the wavelength range from 500 to 700 nm at $T=50$ K and incident angles of $35^{\circ}$ and $65^{\circ}$ for TE polarization. Two regions of high reflectance were found around 570 nm and 650 nm at $35^{\circ}$ and $65^{\circ}$, respectively. For a better understanding of the reflectance behaviour, we have displayed in Fig.~\ref{fig4}(c), the spectra as a function of the whole range of incident angles for TE polarization. One important feature found in our results related with the reflection from a finite multilayer is its sensitive response to the angle of incidence of light~\cite{Joannop:2009}, which is observed in the continuous displacement of PBGs to shorter wavelength as the incident angle increases. On the other hand, as the angle of incidence approaches $90^{\circ}$, the reflection coefficient tends to 1, a result in agreement with the fact that a wave that impinges with a right angle moves parallel to the separation surface of the two media, and therefore, its energy is not transmitted through the surface. The results for $35^{\circ}$ and $65^{\circ}$ (magenta vertical lines) are indicated in the figure, whose bands match perfectly well with measured spectra (Fig.~\ref{fig4} a and b). The gap map also allows to evidence the existence of a third transmission band for wavelengths above 700 nm.  These results allow us to conclude that there exist wavelength ranges where  the radiation does not pass through the PC under any incident angle, results that can be applied, for example, to tune optical transmitter/reflector fabricated below critical superconductor temperature whose response is sensitive to the angle of incidence of light.

For the sake of completeness, we calculate the optical response of $(\text{B}\text{T}\text{O}/\text{Y}\text{B}\text{C}\text{O})_{5}/\text{S}\text{T}\text{O}$ 1D PC when the $\text{Y}\text{B}\text{C}\text{O}$ is in the non-superconducting state, above $T_{c}$.
In this case, the dielectric function is described by the Drude model of a metal given by equation \eqref{eq5}. The steady-state $dc$ conductivity ($\sigma_{0}$) and $dc$ resistivity ($\rho_{0}$) are functions of the plasma frequency $\omega_{p}$ and the damping $\gamma$, and they are related as follows\cite{Orlando:1991}
 \begin{equation}
\label{eq6}
\sigma_{0} = \frac{1}{\rho_{0}}=\frac{nq^{2}\tau}{m}=\frac{\epsilon_{0}{\omega_{p}}^{2}}{\gamma}.
\end{equation} 

Figure ~\ref{fig1}(c) presents the YBCO resistivity measured as a function of the temperature. As it is shown, there is an abrupt change in the resistance of the YBCO. For temperatures below $T_{c}$ there is no resistance, and for temperatures above $T_{c}$ there is a non-zero resistance that varies linearly with the temperature. Resistivity measurements were performed with and without applied laser field radiation on the $(\text{B}\text{T}\text{O}/\text{Y}\text{B}\text{C}\text{O})_{\text{N}}$ 1D PCs. The main idea of this measurement was to confirm that the superconductor ($\text{Y}\text{Ba}_{2}\text{Cu}_{3}\text{O}_{7}$) remains in the superconducting state under applied radiation. Even though the superconductor was radiated at frequencies above its superconductor gap ($\sim$ 8 THz), the 10 mW  laser power in the spectral band of 450-750 nm is not enough to destroy the superconducting state. 

%%%%%%%%%%%%%%%%%%%%%%%%%%%%%%%%%%%%%%%%%%%%%%%
\begin{figure*}[t]
\centering
\includegraphics[width=\linewidth]{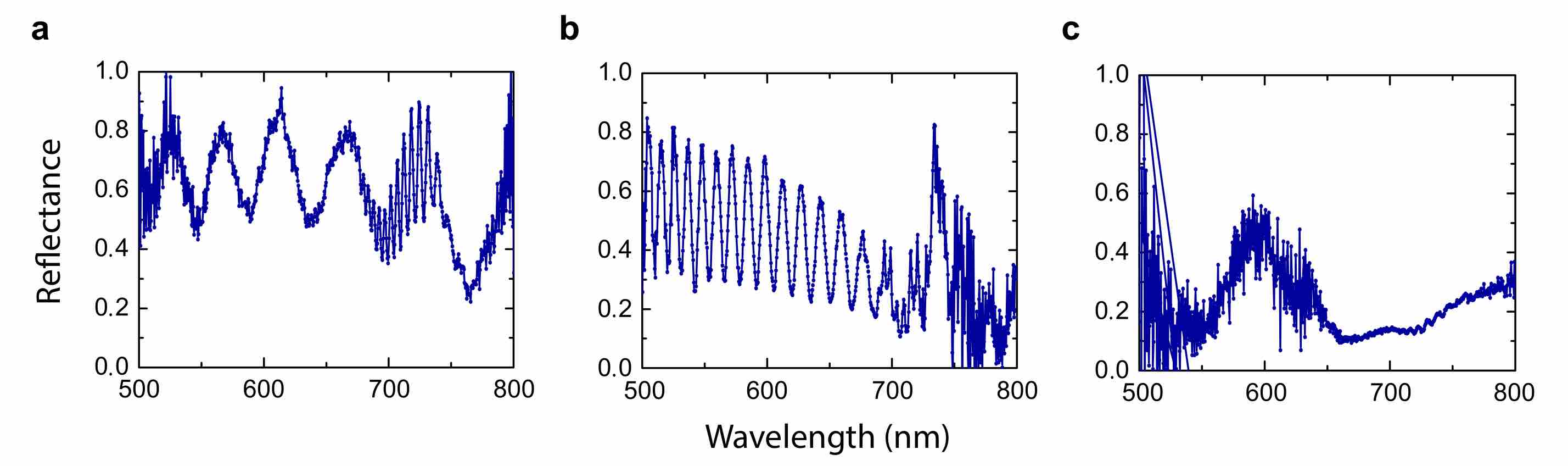}
\caption{Reflectance spectra for $(\text{B}\text{T}\text{O}/\text{Y}\text{B}\text{C}\text{O})_{\text{N}}/\text{S}\text{T}\text{O}$ 1D PCs for periods $N=$ 1 (\textbf{a}), 3 (\textbf{b}), and 5 (\textbf{c}), at $T=20$ K, in the 500 - 800 nm spectral band.}
\label{fig:T20K}
\end{figure*}
%%%%%%%%%%%%%%%%%%%%%%%%%%%%%%%%%%%%%%%%%%%%%%%

Accordingly, in equation \eqref{eq6}, the damping $\gamma$ varies linearly with temperature in accordance with $\rho_{0}=(3.74\times10^{-9})T+(6.90\times10^{-7})$ for $T>T_{c}$, which implies that equation \eqref{eq5} is temperature-dependent. The calculated and measured reflectance are shown in Fig.~\ref{fig4}d, and the behaviour predicted by the Drude model is in excellent agreement with our experimental results about the reflection of electromagnetic waves at the interface between the ferroelectric and the superconductor in normal state. When the electromagnetic wave impinges on the conductor surface induces a conduction current, which leads to a swift damping of the field inside the conductor. This is why metals have excellent reflecting properties (i.e., reflectance close to unity) and are broadly used in mirrors or optical reflectors. According to the above, a high reflectance would be expected in the PC now composed of the superconductor in the normal state. However, as it is observed in the figures~\ref{fig4}d and e, the intensity of the reflectance increases with respect to the previous cases, but it is not total because about 60\% and 40\% of the light is still transmitted through the structure at certain wavelengths between 500 nm and 700 nm, at $T=80$ K and $T=100$ K, respectively. It is important to point out that reflectivity also depends on the thickness of the layers, and our findings clearly illustrate such thickness dependence~\cite{Ian:2016, Shalaev:2019}.

In order to analyse the superelectrons contribution on the optical response of the $(\text{B}\text{T}\text{O}/\text{Y}\text{B}\text{C}\text{O})_{5}/\text{S}\text{T}\text{O}$ 1D PC, in figure~\ref{fig4}f, we report the difference between the reflectance coefficients ($\delta$R) calculated from the subtraction of the reflectance spectra with and without the superelectron effect according to equations \eqref{eq4} and \eqref{eq5}. This result allows us to approximate the magnitude order of the contribution made by the Cooper pairs in the mixed state of the superconductor, below critical temperature, to be of the order $\sim10^{-2}$, and corroborate their significant contribution to the optical response of the superconductor. From the reflectance spectra at $T=100$ K ($T>T_{c}$), and at $T=80$ K, 50 K and 30 K ($T\leq T_{c}$), as can be seen in figure ~\ref{fig3}d, panels D, E, and F, for $N = 5$, we get a decreasing of the reflectance at low temperatures for wavelengths between 500 - 550 nm, and 600 - 650 nm, which implies that the main contribution to the light transmission in the structure is due to the superelectrons. This result highlights the important role played by the superconductor in the optical response of PC. On the other hand, it is worth noting that the Drude model matches perfectly well with the experimental spectrum at 80 K, that is, the temperature around which the states transition occur; however, this model does not adjust the measurements above $T_{c}$ (e.g., at $T=100$ K). Although the Drude model is presented for temperatures above critical temperature, it is evident that for $T>T_{c}$, there exist additional contributions to the electronic properties affecting the optical response. 

Finally, reflectance spectra of $(\text{B}\text{T}\text{O}/\text{Y}\text{B}\text{C}\text{O})_{\text{N}}/\text{S}\text{T}\text{O}$ heterostructures for $N = 1, 3, 5$, at $T=20$ K are shown in Fig.~\ref{fig:T20K}. The reflection spectra for $N=1$ and $N=3$ structures at $T=20$ K clearly shows a constructive interference effect between the incoming light in the YBCO and the reflected light on the YBCO/BTO interface, bearing in mind that the BTO refraction index is larger than that of the YBCO. This is clearly evidenced since the reflected spectra of the $N=3$ structure shows that a maximum (minimum) takes place at each third of the wavelength of the $N=1$ heterostructure. On the other hand, the reflectance spectrum for $N=5$ begins to exhibit the whole reflectance behaviour of the 1D photonic crystal, as reported in~\cite{Gonzalez:2017}.

\section{Conclusions}

We have succeeded at experimentally realising ferroelectric/superconductor 1D photonic crystals as suitable engineered nanosystems for tuning and controlling electromagnetic wave propagation in a wide region of the visible spectrum. We were able to fabricate 1D photonic crystals of $N=1, 3$ and 5 pairs of BTO/YBCO bilayers by DC and RF sputtering, onto polished $\text{SrTiO}_{3}$ (001) substrates, and studied the effects due to temperature and direction of the incident radiation. We have experimentally demonstrated how to tailor the number of PBGs as a function of $N$, and have also been able to quantify and predict, for any $N$,  the frequency range sensitivity and optical properties of the PCs with the direction of the incident EM waves---the larger the angle of incidence the shorter the wavelength and the bigger the width of the PBGs. A key result from an operational point of view is that temperature does not significantly affect the frequency range of the transmission bands, which can be advantageous because this enables either a high or low reflectance, in the whole range of studied temperatures (10-300 K). The contribution made by the Cooper pairs in the mixed state of the superconductor to the PC optical response is of the order of $10^{-2}$: the superelectrons are the most relevant contribution to the light transmission in the structure, at tested wavelengths between 500--550 nm, and 600--650 nm. Finally, and based on the PCs here implemented, several strategies for the development of quantum materials and novel optoelectronic devices, below critical superconductor temperature and different properties of incident light, are proposed. 

\section*{Methods}

\subsection*{Photonic crystal fabrication}
The $(\text{B}\text{T}\text{O}(30~\text{nm})/\text{Y}\text{B}\text{C}\text{O}(73~\text{nm}))_{\text{N=1,3,5}}$ multilayers were grown on a (001) STO substrate by DC/RF sputtering technique at pure oxygen atmosphere and high pressure ($\sim$ 3.0 mBar), with a substrate temperature of $830^{\circ}$C. Power density of 7 W/cm$^{2}$ and 12 W/cm$^{2}$ were used for YBCO and BTO targets, respectively. X-ray diffraction was used to perform structural characterisation using a Co-K${\alpha}$=1.79 $\textup{\AA}$
 wavelength. In addition to the local interface analysis, transmission electron microscopy (TEM) was employed. Cross-sectional lamella was prepared from the deposited thin films using a dual beam FEI Helios Nanolab 600i. Initially, the surface of the films was coated by a platinum layer on a sputtering in order to avoid charge accumulation during the process. Then, two platinum layers were applied to protect and avoid damages on the thin films. The lamella is extracted and submitted to a thinning process which ends with the use of a low energy beam to minimize the amorphization. Finally, the lamella is placed into a TEM grid. A Hitachi TEM-microscope HF3300C was operated at 300 kV to acquire the high resolution images of the cross-sectional lamella. The sample was tilted until the nearest zone axis of STO was parallel to the electron beam. Then, high resolution images were acquired at the different layers and interfaces. In addition, electron diffraction patterns were obtained to identify the crystallographic orientation of the layers respect to the substrate. The electrical properties and the superconductor transition temperature ($T_{C}$) were studied using the traditional four-point technique. The resistance as a function of temperature was measured in the 20-300 K range using silver paint and copper wires. Considering that the top layer was always the BTO layer in the $(\text{B}\text{T}\text{O}/\text{Y}\text{B}\text{C}\text{O})_{\text{N}}/\text{S}\text{T}\text{O}$ multilayer array, it was necessary to analyse the diamagnetism to identify $T_{c}$ through thermal demagnetization measurements using a PPMS Quantum design from 10 to 300 K. 

\subsection*{Wide range laser reflectance measurements}
A supercontinuum fiber laser (Fyla STC 1000) and a monochromator (Fyla TW) were employed to irradiate the sample in the 400-800 nm spectral range. The beam path crosses through a telescope arrangement with a magnification factor of 1x to collimate the laser and control the beam divergence.  The light is addressed with aluminum mirrors (95\% of reflectance) perpendicularly towards the cryostat (Cryostat Advanced Research System) where the sample is placed under a vacuum of $10^{-5}$ bar and a controlled temperature down to 10 K. The reflection angle is controlled with a homemade sample copper holder inside the cryostat, and set to 35$^{\circ}$ and 65$^{\circ}$ for taking different set of measurements. The reflecting path is collected with a collimation lens and coupled to a multimode optical quartz fiber of 400 $\mu$m diameter. The fiber is connected to a spectrometer (HD 4000, Ocean Optics), which finally displays the spectra in a workstation. The reference spectra for the reflectance calculations were taken at high temperatures for which the reflection of the sample is higher than all the others.

\subsection*{1D photonic crystal transfer matrix calculation}

The studied 1D photonic superlattice has a period $d$, and is composed of alternating layers of a dielectric material $\text{Ba}\text{Ti}\text{O}_{3}$ and a superconductor $\text{Y}\text{Ba}_{2}\text{Cu}_{3}\text{O}_{7}$, whose widths are labeled as $a$ and $b$, respectively. The propagation of an in-plane linearly polarized electromagnetic field is of the form $\vec{E}\left(z,t\right)=E\left(z\right)e^{-i\omega{t}}\hat{x}$, along the z-axis (see Fig.~\ref{fig1}a). By using Maxwell'€™s equation for linear and isotropic media, it is demonstrated that the amplitude of the electric field $E(z)$ satisfies~\cite{Gonzalez:2017,Cavalcanti:2007}
%%%%%%%%%%%
\begin{equation}
\label{eq7}
\frac{d}{d{z}}\left[\frac{1}{\textit{n}\left(z\right)\textit{Z}\left(z\right)}\frac{\textit{dE}\left(z\right)}{\textit{dz}}\right]= -\frac{n\left(z\right)}{Z\left(z\right)}\frac{\omega^{2}}{c^{2}}E\left(z\right),
\end{equation}
%%%%%%%%%%%%%%
where $c$ is the vacuum speed of light,  $\textit{n}\left(z\right)= \sqrt{\epsilon\left(z\right)}\sqrt{\mu\left(z\right)}$ and $\textit{Z}\left(z\right)
=\sqrt{\mu\left(z\right)}/\sqrt{\epsilon\left(z\right)}$ are respectively, the refraction index and the impedance of each layer material.
For a photonic crystal composed of alternating layers of two different materials, equation \eqref{eq7} must be solved by assuming both, the electric field and its first derivative continuous across an interface, which means that the two-component function $\psi\left(z\right)= \begin{pmatrix} \textit{E}_{z}\\ \frac{1}{nZ}\frac{dE}{dz}\end{pmatrix}$ is continuous through the photonic structure. This condition may be conveniently written by means of a transfer matrix as $\psi\left(z\right)= M_{i}\left(z-z_{0}\right)\psi\left(z_{0}\right)$, where
%%%%%%%%%%%%%%
\begin{equation}
\label{eq8}
 M_{i}\left(z\right)=
\begin{pmatrix}\cos\left(\frac{\omega\left|n_{i}\right|}{c}z\right)&\frac{n_{i}}{\left|n_{i}\right|}\frac{cZ_{i}}{\omega}\sin\left(\frac{\omega\left|n_{i}\right|}{c}z\right)\\\\-\frac{\left|n_{i}\right|}{n_{i}}\frac{\omega}{cZ_{i}}\sin\left(\frac{\omega\left|n_{i}\right|}{c}z\right)&\cos\left(\frac{\omega\left|n_{i}\right|}{c}z\right)\end{pmatrix}
\end{equation}
%%%%%%%%%%%%%%%%
in such a way that $\psi\left(\pm\frac{a+b}{2}\right)$ may be written as
%%%%%%%%%%%%%
\begin{equation}
\label{eq9}
\psi\left(\pm\frac{a+b}{2}\right)= M_{T}\left(\pm {a},\pm {b}\right)\psi\left(0\right),
\end{equation}
%%%%%%%%%%%%%
with
\begin{equation}
\label{eq10}
M_{T}\left(\pm {a},\pm {b}\right)=M_{2}\left(\pm \frac{b}{2}\right)M_{1}\left(\pm \frac{a}{2}\right)=\begin{pmatrix}P& \pm {Q}\\ \pm {R}&S\end{pmatrix}.
\end{equation}
%%%%%%%%%%%%%%
The reflection coefficients are calculated by 
%\eqref{eq11}
\begin{equation}
R_{N} = \left|\frac{M_{21}}{M_{22}}\right|^{2},
\label{eq11}
\end{equation}
%%%%%%%%%%%
considering the corresponding transfer matrix for $N$ periods, which can be written as~\cite{Markos:2008}
%%%%%%%%%%%
\begin{equation}
\label{eq12}
 M_{N}=
\begin{pmatrix}P U_{N-1}- U_{N-2}& Q U_{N-1}\\\\R U_{N-1} &S U_{N-1}- U_{N-2}\end{pmatrix},
\end{equation}
%%%%%%%%%%%
where $P$, $Q$, $R$ and $S$ are the elements of the transfer matrix for $N=1$. Here, $U_{N}=U_{N}(q)=\frac{\sin\left((N+1)qd\right)}{\sin{qd}}$ are the second-order Chebyshev polynomials.

\bibliography{references}

\section*{Acknowledgements}

This work has been supported by COLCIENCIAS under grant No.~1106-712-49884, the Colombian Science, Technology and Innovation Fund-General Royalties System (Fondo CTeI-Sistema General de Regal\'ias) under contract BPIN 2013000100007, Universidad del Valle CI~71062, and the Center of Excellence for Novel Materials-CENM. L.E.G. is grateful to COLCIENCIAS for financial support (grant No. 811-2018) and Universidad de Ibagu\'e CI~19-505-INT. The authors are grateful to M. A. Hern\'andez-Landaverde of CINVESTAV Queretano-M\'exico, for technical assistance and analysis of XRD measurements, and to J.~Douin, for facilitating TEM measurements at the Centre d'\'E‰laboration de Mat\'eriaux et d'Etudes Structurales (CEMES), Toulouse, France.

\section*{Author contributions statement}

L. E. G. and J. E. O. contributed equally to this work. J.E.O. and E.M. fabricated and structurally characterised the samples, C.A.M-L. conceived and performed the optical measurements together with J.E.O. and E.M., D.R. carried out the TEM characterisation, and L.E.G. performed the analytical and numerical calculations.  L.E.G., J.H.R., J.E.O., C.A.M-L., G.Z., N.P-M., M.E.G. and J.C.G., contributed to the analysis and writing of the manuscript, and with discussion, corrections, feedback and comments.

\end{document}